\documentclass[twocolumn,prl,amsmath,amssymb,unsortedaddress,superscriptaddress,showpacs]{revtex4}
\usepackage{latexsym,epsfig,graphicx}
\usepackage{dcolumn}
\usepackage{bm}
\usepackage{color} 
\usepackage{graphicx}
\usepackage{subfigure}
\usepackage{tikz}
\usetikzlibrary{matrix}

\begin{document}

\title{The escape physics of single shot measurement of flux qubit with dcSQUID}

\affiliation{Department of Physics, University of Illinois at
Urbana-Champaign, Urbana, IL 61801, USA}

\author{Mao-Chuang Yeh}
\affiliation{Department of Physics, University of Illinois at
Urbana-Champaign, Urbana, IL 61801, USA}
\author{Anthony J. Leggett} \affiliation{Department of Physics, University of
Illinois at Urbana-Champaign, Urbana, IL 61801, USA}
\date{\today}

\begin{abstract}

    In most experiments on flux qubits,the "measurement" is performed by
coupling the system to a dc SQUID and recording the distribution of
switching currents for the latter;this measurement protocol is very
far from the classic von Neumann ("projective") scheme,in that very little
information is obtained from a single run,rather one has to repeat the
experiment tens of thosands of times to extract anything useful.Here,
concentrating on the equlibrium behavior of the flux qubit,we carry out an
analytic calculation of the dc-SQUID switching current distribution as a
function of external bias flux on the qubit,and compare our predictions with
the data from experiments conducted at Delft and NTT.

\end{abstract}

\maketitle

    Superconducting rings interrupted by one or more Josephson junctions
("flux qubits") are of great interest both in the context of the extrapolation
of the predictions of quantum mechanics towards the macroscopic level~\cite{LG} and
as possible elements in a future quantum computer~\cite{Clarke}, and
starting with the pioneering experiments of Ref.~\cite{Mooij1,Mooij2,Friedman}  both their
statics and their dynamics have been the subject of extensive experimental
investigations over the last dozen years or so; see for example~\cite{Clarke}. In these experiments the
fundamental quantity whose behavior is of interest is usually the total
magnetic flux, a sum of external and self-induced terms, which threads the
supeconducting loop, so the question arises, how to measure this quantity?
In most experiments to date this has been done by coupling the flux qubit
inductively to a dcSQUID~\cite{Barone}, and detecting the
effect on the rate of escape of the latter from the zero-voltage state.
However, the latter phenomenon, whether occurring as a result of classical(Arrhenius-Kramers) thermal activation or of quantum tunnelling, is a stochastic process; if in the absence of coupling to the qubit one plots the
probability of escape as a function of the bias current applied to the dc
SQUID, one gets a histogram whose width, while generally decreasing
with temperature, is nonzero even in the limit T=0~\cite{Mooij1}. Moreover, in practice the coupling to the qubit is sufficiently weak (cf.below) that its effect on any individual run is very small
compared to the original inherent stochasticity, { so that on any individual run it is impossible to read off the qubit state unambiguiously from the value of the switching current,}
and thus one typically needs to repeat the experiment tens of thousands of times to obtain any useful
information on the behavior of the qubit {  in the ground state.} Such a "measurement" protocol is very far from the classic von Neumann ("projective") scheme, indeed it is closer in concept to the "weak measurement" scheme of
Albert et al.~\cite{Albert}(though no postselection is involved). The situation is further
complicated by the fact that when one inserts realistic values of the relevant
experimental parameters one finds that the coupling to the dc SQUID, while as
mentioned above a small perturbation to the dynamics of the latter, is not
necessarily a "small" perturbation to the behavior of the qubit itself. Thus a
detailed analysis of the measurement process would seem to be of interest.

    Actually, in comparison with the many studies of the intrinsic dynamics of
the qubit itself, there have been to date only a few considertaions in the
literature~\cite{alec, NTTT,NTTT2}of the readout process, and all of these have been
numerical. In particular, the NTT group~\cite{NTTT,NTTT2} have analysed the behavior both in
the absence and in the presence of strong decoherence on the qubit, and concluded
that a von-Neumann-like projection takes place in a regime of strong
qubit-SQUID coupling when the decoherence is sufficiiently strong.

    The goal of the present paper is to give an analytical treatment of the
measurement process for a flux qubit { weakly} coupled to a dc SQUID { ( by "weakly coupled" we mean that the coupling energy is much smaller than the transition energy between the lowest two states of the dcSQUID )}. {We will, initially treat those experiments which probe the static
behavior, that is the dependence(which may be hysteretic) of
the total trapped flux throught the qubit loop on the externally applied
flux~\cite{Mooij1,NTT,NTT2}, and then generalize it to  the analysis of dynamic ones which show the switching probability change due to qubit evolution~\cite{Mooij2} }; thus, the  basic question is how, in the various experimentally interesting
regimes of the relevant parameters, do we expect the switching current
distributions to depend on external flux? { The only two approximations} we shall make(apart from the standard "two-state" projection of the qubit
states) {are   $g\ll 1$(, where $g$ is the dimensionless qubit-dcSQUID coupling parameter defined in the following section ) and} that the zero-point oscillation energy of the dc SQUID is large
compared to all other relevant energy scales of the problem; as we will see
below, this condition has been relatively well satisfied in
existing experiments. 

    In the next section we set up the problem, introduce our notation and
estimate the order of magnitude of the relevant experimental parameters. 
In section 3 we use the above inequality to justify a "harmonic" approximation
to the escape dynamics of the coupled qubit-SQUID syetem, and compare the result
with the data from the NTT and Delft experiments. Sections 4 and 5 are a discussion and summary respectively. { In appendix B, we confirm the
results of section 3 by an explicit calculation of the energy levels of the
coupled system.  } Throughout we work at zero temperature; possible thermal-activation corrections to the zero-temperature
 WKB tunnelling exponents are estimated at the end of section 2 and shown to be small for existing experiments. 
 
 { In static experiments~\cite{Mooij1,NTT,NTT2}, the bias current $I_b$ usually increases {  over a time interval long compared to the inverse of the tunneling rate, so} that we can consider the dcSQUID potential changes adiabatically; 
 on each run, as the bias current $I_b$ across the dcSQUID is increased towards its critical
 value, the value of $I_b$ at which a nonzero voltage drop develops (the "switching current" $I_{SW}$) is recorded, and it is the distribution of $I_{SW}$ averaged over many runs which constitutes the raw data of the experiment. In one experiment of this type~\cite{Mooij1} the Delft group determined the average value of $I_{SW}$ as a function of the flux applied to the qubit, which determines the groundstate energy and wave function of the latter. In a  second such experiment, the NTT group\cite{NTT2} found a pattern of two peaks in the distribution of $I_{SW}$ which cross as a function of the applied flux( the so-called $\chi$-structure, the fig(4a) of Ref.~\cite{NTT2}), and interpreted these as corresponding to the switching behavior for the ground state and excited state of the qubit.}{

    In the dynamic{( Rabi oscillations or Ramsey fringe) experiments}~\cite{Mooij2}, the ramping rate of $I_b$ is faster than in the static measurement and $I_b$ usually reaches  a definite value at which the averaged switching probability is maintained at $50\%$. The bias current $I_b$ usually consists of a short pulse followed by a  trailing plateau, where the height of short pulse is just equal to the required value for $50\%$ switching probability and the height of the trailing plateau is about $70\%$ of short pulse{~\cite{plateau}.} The purpose of the trailing plateau is to avoid missing the voltage signal due to the retrapping of SQUID phase.
    
      { In a Rabi experiment,} the qubit is initialized to the ground state and then manipulated coherently beteen the ground state and excited state by applying microwaves of a frequency equal to the energy difference of the qubit states. { To obtain the switching probability, we repeat the switching-event detection, applying the current bias pulse right after the microwave operation on qubit. The switching probability is observed to oscillate as a function of the microwave operation time; this is the Rabi oscillation phenomenon.}    
     
      Similarly, Ramsey interference is obtained by measuring the switching probability right after the qubit has been manipulated by two $\frac{\pi}{2}$  pulses separated by a varying delay time, with the microwave frequency detuned from resonance by $\delta F$;  the Ramsey fringe  period is then $\frac{1}{\delta F}$. {The first pulse creates an equal superposition of the qubit energy states, which then precess at different rates, and the second one is applied to give different occupancy between qubit energy states; in general the ratio of occupancy oscillates and depends on the delay time between the pulses.}}

\section{The basic analysis of qubit-SQUID coupled system}

Let's first introduce our starting model. {It is believed that 
we can give a simplified description of the flux qubit  as a} two-level
system with the two flux states $\{\left| R \right\rangle , \left|
L \right\rangle \}$ and a tunneling energy $\Delta$ between them. The qubit effective Hamiltonian can be represented by the
Pauli spin matrices $\sigma_{z,x}$, that is
\begin{eqnarray}
H_{q} = \varepsilon \sigma _z  - \Delta \sigma _x 
\label{eqn:Hq}
\end{eqnarray}
with $2\varepsilon$ being the energy difference between two flux
states. The relation between $\varepsilon$ and the flux
$\Phi_q$ applied to the qubit is~\cite{Mooij2,NTTT}
{
\begin{eqnarray}
\varepsilon \equiv\varepsilon (f_q){\approx I_p\Phi_0\left( {f _q - \frac{1}{2} }
\right),}
 \label{eqn:Eq}
\end{eqnarray}
where $I_p$ is the maximum qubit persistent current~\cite{three}} and we have defined $ f_q  \equiv {{\Phi _q }
\mathord{\left/
 {\vphantom {{\Phi _q } {\Phi _0 }}} \right.
 \kern-\nulldelimiterspace} {\Phi _0 }}$ and
$f_{SQ}  \equiv {{\Phi _{SQ} } \mathord{\left/
 {\vphantom {{\Phi _{SQ} } {\Phi _0 }}} \right.
 \kern-\nulldelimiterspace} {\Phi _0 }}
$ for flux parameters of the qubit and dc-SQUID respectively, where
 $\Phi_q$ and $\Phi_{SQ}$ are the corresponding applied fluxes and $\Phi _0 =h/2e $ is the flux quantum.

  On the other hand, we can consider the dc-SQUID as a system of one
degree of freedom $x$ {which is the average of the
phases of the two junctions~\cite{SQUID DOF}}. As we have mentioned, the SQUID potential
changes with the applied flux $\Phi _{SQ}$ and bias current
$I_b^{} \left( t \right)$, and each qubit state induces a different
flux on the SQUID. The total flux on the SQUID will be $ \Phi
_{SQ} \pm\Phi _M $ depending on which of the qubit states is realized,
where $2\Phi _{\rm{M}}$ is the net flux difference on the SQUID
induced by the qubit states. {Denoting by $E_{J0}$ the Josephson
energy of a single junction,} we can construct our SQUID phase
potential, including the effect of the induced flux from the qubit, as follows: 
\begin{eqnarray}
 U_{0}\left({x,f_{SQ}+ g\sigma _z }\right)
 &=&-2E_{J0} \cos \left[ {\pi f_{SQ}  +\pi
g\sigma _z } \right]\cos \left[ x \right]\nonumber\\
 &-& \left( {\frac{\hbar
}{{2e}}} \right)I_b^{} \left( t \right)x
 \label{eqn:STU}
\end{eqnarray}
{, where we define the dimensionless coupling(or flux) parameter $ g$ as $\frac{{\Phi_{\rm{M}} }}{{\Phi _0 }}$. 
 Expanding in terms of the small parameter $g$, we can rewrite the potential as}
\begin{eqnarray}
U_{0}\left({x, f_{SQ}+ g\sigma _z }\right) \approx U_0^{}
\left( {x, f_{SQ} } \right)+\varepsilon _{{\mathop{\rm int}} }
\left( x \right)\sigma _z   \label{eqn:Utot}
\end{eqnarray}  
  with the coupling energy $\varepsilon_{\rm int}\left( x \right)$
 defined by the formula
\begin{eqnarray}
 \varepsilon _{{\mathop{\rm int}} } \left( x \right)=2\pi
gE_{J0}\sin \left[ {\pi f_{SQ} } \right]\cos \left[ x \right]. \label{eqn:vxint}
\end{eqnarray} 
{In the above approximation we have assumed  $g\ll 1$, which is additional to the weak coupling assumption that the coupling energy is much smaller than the ground state energy of the dcSQUID.} 
 
  Considering the potential of Eq.(\ref{eqn:Utot}) together with the SQUID kinetic 
 energy and the qubit Hamiltonian, the total Hamiltonian of the coupled system is
\begin{eqnarray}
H = H_q  + H_{SQ}  + H_{coupling} \label{eqn:totH}
\end{eqnarray}
, where we have
\begin{eqnarray}
 H_{\rm SQ}  = \frac{{ - \hbar ^2 }}{{2m}}\partial _x ^2  - 2E_{J0} \cos \left[ {\pi f_{SQ} } \right]\cos \left[ x  \right] -\left( {\frac{\hbar }{{2e}}} \right)I_b x  \label{eqn:HSQ}
\end{eqnarray}
\begin{eqnarray}
 H_{\rm coupling}= \varepsilon _{{\mathop{\rm int}} } \left( x \right)\sigma _z.\label{eqn:Hcoupling}
\end{eqnarray}

Here the effective mass $m\equiv  2C_0 \left( {\frac{\hbar }{{2e}}} \right)^2$,
and $C_0$ is the capacitance of one junction of the SQUID.

    Before appling the WKB decay analysis to our coupled system, { we need to make a further approximation to the potential. With a Taylor expansion} around
     the minimum of the well $x=x_0$, where $ \sin \left[ {x_{\rm{0}} } \right] 
     = \frac{{\left( {\frac{\hbar}{{2e}}} \right)I_b^{} \left( t \right)}}{{2E_{J0} \cos \left[ {\pi f_{SQ} } \right]}}$, we can approximate the washboard
    potential $U_{0}\left({x, f_{SQ} }\right)$ as
\begin{eqnarray}
v\left( R \right) = \frac{1}{2}kR^2  - \frac{1}{2}\frac{k}{{R_c
}}R^3. \label{eqn:APXU}
\end{eqnarray}

  Here we have $R=x-x_0$ and  $ k = 2E_{J0} \cos \left[ {\pi f_{SQ}
} \right]\cos \left[ {x_0 } \right]$. The classical turning point
is defined by {$R=R_c=3\cot \left[ {x_0 } \right]$~\cite{turning};} that is
$v\left( R \right)>0$ for $0<R<R_c$. 
 Besides, we can also approximate the form of
 $\varepsilon_{\rm int}\left( x \right)$ around $x_0$, that is
 
{\begin{widetext}
\begin{eqnarray}
{\varepsilon }_{int} \left( R \right)
 =\pi g\tan \left[ {\pi f_{SQ}
} \right]\left( {  k - 3\frac{k}{{R_c }}R - \frac{k}{2}R^2 +
\frac{1}{2}\frac{k}{{R_c }}R^3 } \right). \label{eqn:vint}
\end{eqnarray}
\end{widetext}}

{ Finally, we obtain our approximate Hamiltonian }
\begin{eqnarray}
H_{approx}=
\frac{{ - \hbar ^2 }}{{2m}}\partial _R^2 + v\left( R \right)
+\varepsilon\left( R \right)\sigma _z  - \Delta \sigma _x \label{eqn:APXH}
\end{eqnarray} with the qubit total bias energy 
$ \varepsilon\left( R\right)\equiv\varepsilon +\varepsilon_{int} \left( R \right)$.
{ The analysis bellow is based on this
 approximate hamiltonian.} The first two terms should determine the
 standard decay physics of SQUID single shot measurement.
 And near the minimum of the well, the effect of the
cubic term of the potential $v\left( R \right)$ is relatively
small { in comparison with } the square term, and therefore we can consider our potential as a simple harmonic one with oscillation frequency $\omega  = \sqrt {\frac{k}{m}}$, where the corresponding simple harmonic energy levels are 
$\left| {\rm{n'}}\right\rangle $. { For the experimental setup of NTT
group~\cite{NTTT}, we have the parameters:

\begin {table}[h] \centering \begin{tabular}{|c|c|}
   \hline 
   {$\pi g 
    $} & $\sim 0.003$  \\ \cline{1-2} 
     $f_{SQ}$ & $0.4$  \\ \cline{1-2} 
     $\Delta$   & $  1{\rm{GHz}}=6.6\times 10^{ -25}  J$  \\ \cline{1-2}
   {$I_{c0}$} & $\sim 200nA $ \\ \cline{1-2} 
    $m\equiv  2C_0 \left( {\frac{\Phi_0 }{{2\pi}}} \right)^2$ & $7.4\times 10^{ - 44} {Js^2 } $ \\ \cline{1-2}
    $E_{J0}=\frac{\Phi_0 I_{C0}}{2\pi }$ & $100GHz= 6.6\times 10^{-23} J$  \\ \cline{1-2} 
    $  k = 2E_{J0} \cos \left[ {\pi f_{SQ}
    } \right]\cos \left[ {x_0 } \right]$ & $ 4.1 \times
       10^{ -23} \cos \left[ {x_0 } \right]J$   \\ \cline{1-2} 
    $\nonumber\omega \hbar=\sqrt{\dfrac{k}{m}}\hbar $ & $2.5\times 10^{ -24} \sqrt {\cos \left[ {x_0 } \right]}  J$   \\ \cline{1-2} 
    \end{tabular} 
    \end{table}




 The corresponding parameters for the Delft experiment~\cite{Mooij1} are:
 
 \begin {table}[h] \centering \begin{tabular}{|c|c|}
    \hline 
     {$\pi g 
      $} & $\sim 0.005$  \\ \cline{1-2} 
       $f_{SQ}$ & $0.76$  \\ \cline{1-2} 
    $\Delta$   & $ 0.33GHz=2.2\times 10^{ -25}  J  $ \\ \cline{1-2}
    {$I_{c0}$} & $\sim 110nA $ \\ \cline{1-2} 
     $m\equiv  2C_0 \left( {\frac{\Phi_0 }{{2\pi}}} \right)^2$ & $1.3\times 10^{ - 46} {Js^2 } $ \\ \cline{1-2}
     $E_{J0}=\frac{\Phi_0 I_{C0}}{2\pi }$ & $ 3.6\times 10^{-23} J$  \\ \cline{1-2} 
     $  k = 2E_{J0} \cos \left[ {\pi f_{SQ}
     } \right]\cos \left[ {x_0 } \right]$ & $ 5.3 \times
        10^{ -23} \cos \left[ {x_0 } \right]J$   \\ \cline{1-2} 
     $\nonumber\omega \hbar=\sqrt{\dfrac{k}{m}}\hbar $ & $6.7\times 10^{ -23} \sqrt {\cos \left[ {x_0 } \right]}  J$   \\ \cline{1-2} 
     \end{tabular} 
     \end{table}


{ Before analyzing the physics of our system, we can briefly
estimate the expectation value of the coupling energy based on the
ground state wave function of the harmonic well, which is
$\left\langle 0' \right| 
\varepsilon _{int} \left( R
\right)\left| 0' \right\rangle\sim 10^{ - 25} J$.} It is of the same
order of magnitude as the qubit tunneling energy, but much smaller
than the zero point energy of the dcSQUID:

{
\begin{eqnarray}
{\begin{array}{*{20}c}
   {{\omega  \hbar }}  \\
     \left({  10^{ - 24} J} \sim {  10^{ - 23} J}\right)   \\
\end{array}}  \gg {\begin{array}{*{20}c}
   \Delta   \\
   \left( \sim { 10^{ - 25} J}\right)   \\
\end{array}}  \cong  {\begin{array}{*{20}c}
   {\left\langle 0' \right|\varepsilon _{\rm{int}} \left( R  \right)\left| 0' \right\rangle }
   \\
   \left( { \sim 10^{ - 25} J}\right) . \nonumber
\end{array}}
\end{eqnarray}}
{In comparison with these energy scales, the estimated  dc SQUID escape
rate(   $\Gamma \hbar\sim 10^{ - 26} J $  for the bias current $I_b$ around the value which gives the maximun switching probability of dcSQUID  ) has the smallest energy among them. Here we estimated  $\Gamma \hbar$ by the formula }
\begin{eqnarray}
\Gamma ^{}  = \omega 60^{1/2} \left( {\frac{{B }}{{2\pi \hbar }}} \right)^{1/2}
\exp - [\left(B /\hbar\right)\left(1+\frac{0.87}{Q} \right) ].\label{eqn:Gamma}
\end{eqnarray}, 
where ${\rm{B = }}\frac{{\rm{8}}}{{{\rm{15}}}}m\omega R_c^2$
 is the decay bounce action and $Q=\omega R_s C_0$ is the dampingfactor with net resistance $R_s$~\cite{WKB}.
 { As a result, we can ignore the escape effect in some steps by considering its relative small energy scale, which can help us to simplify the analysis of qubit-SQUID system in the next section.{ We note in passing that }the thermally activated escape rate $\Gamma_{T}\hbar$~\cite{thermal} is of order $10^{ - 55} J $, which is negligibly small. Thus, it is adequate to consider only zero-temperature WKB tunnelling in our analysis.} 

\section{The Harmonic approximation for decay dynamics of qubit-SQUID system}

  Because the decay dynamics is the smallest energy scale in our
  system, the behavior of the harmonic kernel is believed to
 dominate most of the properties of the wave function of the coupled system.
 It will simplify the analysis if we can ignore the
   decay mechanics for a moment. What we want to do is to try to approximately diagonalize the system {in the ground state} (at least within the harmonic region of potential)
    before we really take the decay physics into account.

 Therefore, the first step is to to drop the cubic terms of the potential:
\begin{eqnarray}
v\left( R \right) &{\to}& v^H \left( R \right)=\frac{1}{2}kR^2\nonumber\\
\varepsilon \left( R \right) &{\to}& \varepsilon ^H\left( R
\right) =\varepsilon+ \pi g\tan \left[ {\pi f_{SQ} } \right]\left( { 
k - 3\frac{k}{{R_c }}R - \frac{k}{2}R^2 } \right)\nonumber
\end{eqnarray}
{ Then we can get a new simple-harmonic-approximate Hamiltonian, namely}
\begin{eqnarray} 
H^H &=& \frac{{ - \hbar ^2 }}{{2m}}\partial _R^2 + v^H\left( R \right)
+\varepsilon^H \left( R \right) \sigma _z  - \Delta \sigma _x.
 \label{eqn:APXHH}
\end{eqnarray}
  This Hamiltonian will be helpful for us to determine the system's behavior within the harmonic region of the potential.

 According to our previous analysis of the energy scales, the ground
 state expectation value of the last two terms in $H^H$ is much
 smaller than the ground state energy of the harmonic well.
 Therefore, we may treat these two terms by a perturbation analysis.
  On the other hand, we have assumed that the system always starts
 from the lowest simple harmonic state of the well, in order to
 follow the traditional WKB decay analysis.

 Once we rewrite our hamiltonian in harmonic form {$H^H$, the next step is to
 change our representation into the new spin basis that is determined by diagonalization of the ground state expectation value of perturbation term }
 $\left\langle 0'
\right|\varepsilon\left( R \right)\sigma _z - \Delta \sigma _x
\left| 0' \right\rangle${, where $\left| 0' \right\rangle$, as we defined in last section, is the simple harmonic ground state at minimum of the SQUID potential.} The relations between new and old Pauli
matrix are { $ \tau _z  = \cos \chi  \sigma _z  + \sin \chi
\sigma _x$ and $\tau_x  = -\sin \chi \sigma _z  + \cos \chi
\sigma _x $,} where the angle is defined by 
\begin{eqnarray}
\sin \chi = \frac{{
- \Delta }}{{\sqrt {\varepsilon _{00}^{H2}  + \Delta ^2 } }}{\rm{
}}, \ \cos \chi = \frac{{\varepsilon _{00}^H }}{{\sqrt
{\varepsilon _{00}^{H2}  + \Delta ^2 } }}\label{eqn:angle}
\end{eqnarray}   with the definition  $
\varepsilon _{00}^H  \equiv \left\langle 0' \right|\varepsilon^H
\left( R \right)\left| 0' \right\rangle 
. $

 After rewriting our hamiltonian in the new spin basis, we can
rearrange it in the following form:
\begin{eqnarray}
H^H  &=&  {\frac{{ - \hbar ^2 }}{{2m}}\partial _R^2  + v^H \left(
R \right) + \left( {\varepsilon ^H \left( R \right)\cos \chi
  - \Delta \sin \chi  } \right)\tau _z
} \nonumber\\
 &{}&- \left( {\varepsilon ^H \left( R \right)\sin \chi
 + \Delta \cos \chi } \right)\tau _x
 \label{eqn:APXHH1}
\end{eqnarray}

In this new representation, the Hamiltonian can be divided into
two parts; one is the off diagonal part $V^H  =  - \left( {\varepsilon ^H \left( {R}\right)\sin \chi   + \Delta \cos \chi } \right)\tau _x$, and the rest is the diagonal part $H_d^H$. {  The
diagonal part describes the physics of two independent harmonic
channels with different spring constants; their eigen states are
denoted as $\left| {n'_ -  } \right\rangle$ and $ \left|
{n'_+  } \right\rangle $. The off diagonal part now can be
considered as new perturbation term instead, and its perturbative
correction to the eigen energy of two harmonic channels can be
evaluated with $2\delta\omega \hbar$ being the energy difference between two states $\left| {0'_ -   } \right\rangle$ and $\left| {0'_+   } \right\rangle$, where the dominate term is
 $ 
  \frac{{\left| {\left\langle {0'_ +
 } \right|V^H \left| {0'_ -   } \right\rangle }
\right|^2 }}{{\left[ {2\delta\omega \hbar} \right]}} \cong \left(
{ \omega \hbar } \right)\frac{{g^5 }}{{1024}}\tan ^5
 \left[ {\pi f_{SQ} } \right]\cos\chi\sin ^2 2\chi \sim 10^{ - 34} J
$, see appendix A. Threrefore, this perturbation correction to the eigen energy is much smaller than the decay energy scale of the SQUID near the maximum of switching probability,  so that we will neglect this off diagonal term in the following discussion. }

  The last step is to restore the cubic terms in the SQUID potential
 and the coupling energy, that is to replace $v^H \left(
R \right)$ and $\varepsilon ^H \left( R \right)$ by $v \left( R
\right)$ and $\varepsilon \left( R \right)$ respectively in the
diagonal part of $H^H$. Finally, we have the kernel Hamiltonian
which describe the dominant physics of the system:
\begin{eqnarray}
H_d  &=& \frac{{ - \hbar ^2 }}{{2m}}\partial _R^2  + v\left( R
\right) + \left( {\varepsilon \left( R \right)\cos \chi -
\Delta \sin \chi  } \right)\tau _z\nonumber\\
 &=& \left( {\begin{array}{*{20}c}
   {H_ + ^{} } & 0  \\
   0 & {H_ - ^{} }  \\
\end{array}} \right)\label{eqn:KERH}
\end{eqnarray}

, here we have $ H_ \pm ^{}  = \frac{{ - \hbar ^2 }}{{2m}}\partial
_{R}^2  + k_0^ \pm   + \frac{1}{2}\overline k _ \pm ^{} R^2  -
\frac{1}{2}\frac{{\overline k _ \pm ^{} }}{{\overline R _{c\pm}
}}R^3 $ and parameters given by:
\begin{eqnarray}
\begin{array}{l}
 \overline k _ \pm ^{}  \equiv k\left( {1 \pm \pi g\cos \chi  \tan \left[ {\pi f_{SQ} } \right]\left( {1 + \left( {\frac{3}{{R_c^{} }}} \right)^2 } \right)} \right) \\
 \overline R _{c\pm}  \equiv R_c \left( {1 \pm \left( {\frac{3}{{R_c^{} }}} \right)^2 \pi g\cos \chi\tan \left[ {\pi f_{SQ} } \right]} \right) \\
 k_0^\pm =\mp \left( \sqrt {\varepsilon _{00}^{H2}  + \Delta ^2}+\frac{\pi g}{4}\omega \hbar \tan \left[ {\pi f_{SQ} } \right]\cos \chi 
\right).\label{eqn:Para}
 \end{array}
\end{eqnarray}
{
Based on Eq.(\ref{eqn:KERH}), we can clearly see that $H_+$ and $H_-$ describe two independent decay chanels (for spin$+$ and spin $-$) respectively. Therefore, we simplify the escape dynamics of qubit-dcSQUID composite system by the two-channel decay dynamics where each channel has its own ground state and the corresponding conventional escape rate (as shown in Fig. \ref{fig:f03}); the composite system can be in superposition of these ground states.}

\begin{figure}[t]
\centering
   \includegraphics[width=7cm]{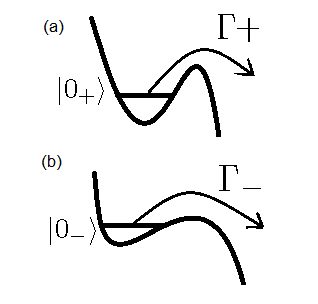}
  \caption{{ The two possible escape ways corresponding to two qubit states $\left| +\right\rangle $  and $\left| -\right\rangle $ are shown in (a) and (b) where  two SQUID potentials have different ground states($\left| 0_+\right\rangle $  and $\left| 0_-\right\rangle $) and independent escape rates ($\Gamma+$ and $\Gamma_-$).  }}  
  \label{fig:f03}
\end{figure}

 The corresponding oscillation frequency and bounce actions
are
\begin{eqnarray}
\omega _ \pm   = \sqrt {\frac{{\bar k_ \pm  }}{m}}  \simeq \omega
  \pm \delta \omega.\label{eqn:omegapm}
\label{eqn:omgpm}
\end{eqnarray}
\begin{eqnarray}
B_ \pm   = B  \cdot \left( {\frac{{\omega _ \pm  \overline R _{c
\pm }^2 }}{{\omega  R_c^2 }}} \right)
 \label{eqn:Bpm}
\end{eqnarray}
 Here we have 
 \begin{eqnarray}
 \delta \omega  \simeq  \frac{\pi g}{2} \omega \cos \chi
 \tan \left[ {\pi f_{SQ} } \right]\left( {1 + \left(
 {\frac{3}{{R_c^{} }}} \right)^2 } \right) \label{eqn:deltaomega}
 \end{eqnarray}.

 With the use of Eq.(\ref{eqn:Gamma}),(\ref{eqn:omgpm}), and(\ref{eqn:Bpm}), we
 can  easily derive the decay rate of two spin channels:
\begin{eqnarray}
\Gamma _ \pm ^{}
 &=& \omega _ \pm  60^{1/2} \left(
{\frac{{B_ \pm }}{{2\pi \hbar }}} \right)^{1/2} \exp  - \left[
{\left( {{{B_ \pm } \mathord{\left/
 {\vphantom {{B_ \pm  } \hbar }} \right.
 \kern-\nulldelimiterspace} \hbar }} \right)\left( {1 + \frac{{0.87}}{{Q_ \pm  }}} \right)} \right]\nonumber\\
 &=& \Gamma  \left( {\frac{{\omega _ \pm  }}{{\omega }}}
\right)\left( {\frac{{B_ \pm  }}{{B }}} \right)^{1/2} \exp  -
\left[ {\frac{{B }}{\hbar }\left( {\left( {\frac{{\omega _ \pm
\overline R _{c \pm }^2 }}{{\omega  R_c^2 }} - 1} \right)}
\right.} \right.\nonumber\\
 &+& \left. {\left. {\frac{{0.87}}{{Q }}\left( {\frac{{\overline
R _{c \pm }^2 }}{{R_c^2 }} - 1} \right)} \right)} \right]\label{eqn:GammaPM0}
\end{eqnarray}

With the further approximation  $Q\gg 1$, we have
\begin{eqnarray}
\Gamma _ \pm \simeq {\rm{   }}\Gamma  \left(1\pm {\frac{{3 }}{2
}}{\frac{{\delta \omega }}{\omega }}\pm{\frac{{\delta R_c}}{R_c }}
\right)\exp  \mp \frac{B}{\hbar }\left( {\frac{{\delta \omega
}}{\omega }}+2{\frac{{\delta R_c}}{R_c }} \right).
\label{eqn:GammaPM}
\end{eqnarray}
 Here 
 \begin{eqnarray}
 \delta R_c\equiv \pm \left(\overline R
 _c^ \pm-R_c\right)={ {\frac{9\pi g}{{R_c }}}\cos \chi
  \tan \left[ {\pi f_{SQ} } \right]}
 \end{eqnarray}.

   \begin{figure}[t]
 \centering
 \includegraphics[width=7cm]{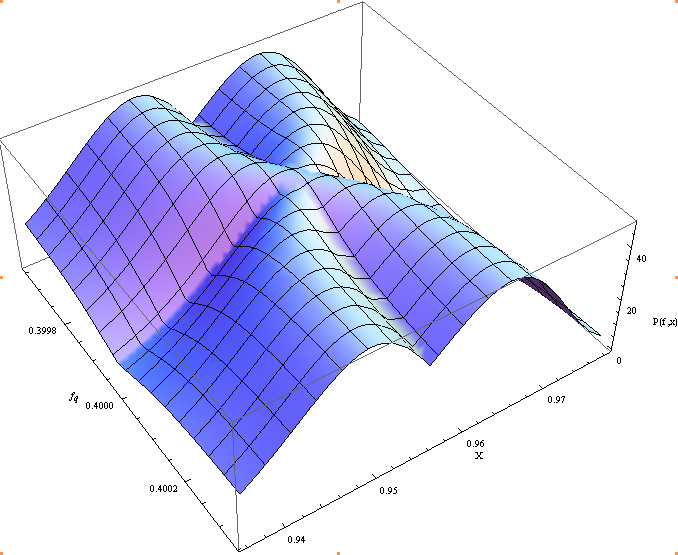}
  \caption{ { The calculated  switching current
  probability corresponding to the ground state and excited state in the NTT group's setup. Here the X-axis and Y-axis
   represent the applied flux $f_{q}$ and current parameter $
x \equiv I_b/{2I_{C0} \cos \pi f} $ respectively. We can see
    two clear {ridges crossing one another, where each ridge structure corresponds to escape-probability behavior for flux qubit being in either the ground state or excited state;} { this $\chi$-cross structure was shown in NTT's experiments(i.e. Fig.4(a) of \cite{NTT2}) with ground and excited states being in thermal distribution.}}
}   \label{fig:f01}
\end{figure}

   Together with the assumption that the state of the
coupled system evolves adiabatically with the change of bias
current, we can use the formula
  \begin{eqnarray}
P_ \pm  \left( y \right) = \frac{{\Gamma _ \pm  \left( y
\right)}}{{dy/dt}}\exp \left[ { - \frac{1}{{dy/dt}}\int_0^y
{\Gamma _ \pm  \left( {y'} \right)dy'} } \right] \label{eqn:swprob}
\end{eqnarray}

  to determine the switching current probabilities corresponding to the ground and excited states of the flux qubit.
 Here $y \equiv \frac{{I_b \left( t
\right)}}{{2I_{C0} \cos \pi f_{SQ} }}$ is the current parameter,
and $I_{C0}  = \frac{{2e}}{\hbar }E_{J0} $ is the critical current
of each junction on the dc-SQUID. We will assume that in static experiments of the type conducted in Refs \cite{NTT2} and \cite{Mooij1} the ramping rate lies in the range  $\sim10^3 - 10^5 Hz$.  The calculated results
for the experiments of the NTT(Fig.(4) in~\cite{NTT2}) and Delft groups~\cite{Mooij1} are shown in the
Fig.\ref{fig:f01} and Fig.3 respectively.
\begin{figure}[t]
\centering
   \includegraphics[width=7cm]{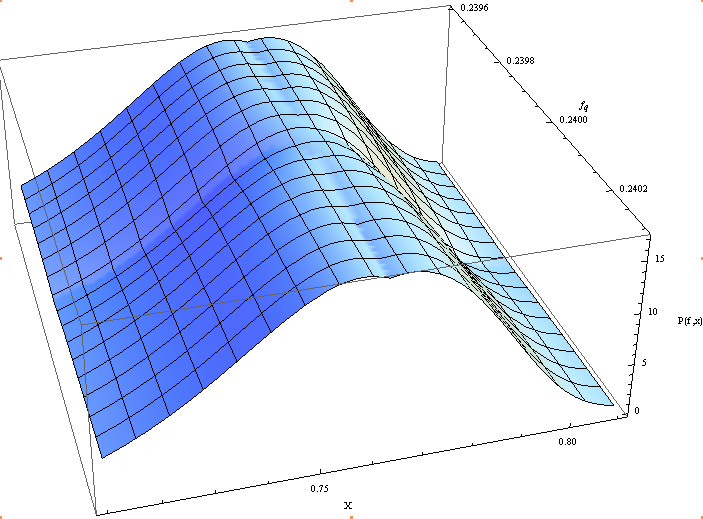}
  \caption{  The switching current probability in the experiment of the Delft group calculated from our formula. Here we can't clearly see
   the flux dependence of the ground and excited states.}
 \label{fig:f02}
\end{figure}
  In the limit where $
{\frac{B}{\hbar }\left( {\frac{{\delta \omega }}{\omega
}}\right) , \frac{B}{\hbar }\left( {\frac{{\delta R_c }}{R_c }}
\right) \ll {\rm{1}}} ${, (a condition satisfied  in most experiments),} we have
the further approximate formula for the decay rate difference
\begin{eqnarray}
 {\frac{{\left| {\Gamma _ +   - \Gamma _ -  } \right|}}{{\Gamma
 }} \simeq 2\left( {\frac{{B }}{\hbar } - \frac{3}{2}}
\right)\frac{{\delta \omega }}{\omega } + 2\left( {\frac{{2B
}}{\hbar } - 1} \right)\frac{{\delta R_c }}{{R_c }}}
\label{eqn:GammaDiff}
\end{eqnarray}
, where $\frac{{B}}{\hbar } - \frac{3}{2}>0$  because of the condition for validity of the WKB analysis. { From this equation, we can see that there are two ways to make the decay rate difference larger( to the point that the switching distributions of the two qubit energy states are distinguishable from one another); one is to have the coupling $g$   large so as to give large ratios $\frac{{\delta \omega }}{\omega }$ and $\frac{{\delta R_c }}{{R_c }}$, which corresponds to pushing the two current distributions apart, and the other one is to make a larger  bounce action $\frac{{ B}}{\hbar } $, which increases the resolution of the measurement by reducing the width of the switching current distribution.} According to the above analysis, if we want the ground state flux-dependent switching current distribution to be easily distinguishable from the excited one, as shown by the crossing feature in FIG.\ref{fig:f01}, a large capacity $C_0$ and Josephson coupling energy $E_{J0}$ is needed to increase $B$ { (since $\frac{{B }}{\hbar } \propto \sqrt
{C_0 E_{J0} \cos \pi f_{SQ}}$.)} {That's why the NTT experiment, with its large values of $C_0$ and $E_{J0}$, can show two distinct current distributions for qubit's eigen states.}

  From Eqns.(20), (23), and(\ref{eqn:GammaDiff}), we know that the escape rate
  difference is also proportional to $ \cos \chi  =\frac{{\varepsilon _{00}^H }}
  {{\sqrt {\varepsilon _{00}^{H2}  +\Delta ^2 } }}$. The explicit formula
  for $\varepsilon _{00}^H$ is
\begin{eqnarray}
\varepsilon _{00}^H  &=& \varepsilon  + \left\langle 0'
\right|\varepsilon_{int} \left( R \right)\left| 0'
\right\rangle\nonumber\\
 &=& { I_p\Phi_0\left( {f _q - \frac{1}{2} }\right) + \pi gk\tan \left[ {\pi f_{SQ} } \right]\left( {  1-\frac{{\omega
 \hbar }}{{4k}}} \right).}
\end{eqnarray}
Here we can easily see that the ground state expectation value of
coupling energy $\left\langle 0' \right|\varepsilon_{int} \left( R
\right)\left| 0' \right\rangle$ induces an effective phase(or flux)
bias on qubit even if the current through the dc SQUID is not turned on.
The estimated flux bias on qubit(for the Delft group) is $ \delta f
_q = \frac{{\pi g k}}{{I_p\Phi_0 }}\tan \left[ {\pi f_{SQ} }
\right]\left( {\frac{{\omega \hbar }}{{4k}} - 1} \right) =
0.0001.$ It is much smaller than what is due to the appearance of
SQUID circulating current right after switching on bias
current~\cite{Mooij2}.
 Thus we can ignore this flux-bias in experiments of Delft and NTT
 group. 

 { Another physics aspect behind} our harmonic-aproximation analysis can be understood as follows. It is well known that a convenient approach to evaluate quantum decay is to utilize the current-density-type formula 

\begin{eqnarray}
\Gamma  = {{J\left( R \right)} \mathord{\left/
 {\vphantom {{J\left( R \right)} {\int_{ - \infty }^R {dR'\left| {\psi \left( {R'} \right)} \right|^2 } }}} \right.
 \kern-\nulldelimiterspace} {\int_{ - \infty }^R {dR'\left| {\psi \left( {R'} \right)} \right|^2 } }}
\end{eqnarray}

 Here we have
 $
 J\left( R \right) = \left( {{\hbar  \mathord{\left/
  {\vphantom {\hbar  m}} \right.
  \kern-\nulldelimiterspace} m}} \right){\mathop{\rm Im}\nolimits} \psi _{out}^* \frac{{\partial \psi _{out} }}{{\partial R}}
 $
   being the outgoing probability current near the turning point $ R_c$ and 
   $   \int_{ - \infty }^R {dR'\left| {\psi \left( {R'} \right)} \right|^2 }  $ as the normalization factor~\cite{WKB}.

   Because the harmonic wave function around the minimum of the potential dominates the outgoing wave function $ \psi _{out} $ near $ R_c$ in the quasiclassical approximation, the quantum decay is totally controlled by the wave function within the harmonic region.  Therefore, once we find an aprropriate basis to effectively diagonalize the harmonic part of the hamiltonian into two escape channels( where the minimized tunneling energy between the channels is negligible comapared to escape energy), we may consider the escape dynamics separately in the two channels.

   A more comprehensive way to evaluate the energy levels of the composite system in a perturbation approximation is given in the appendix B, which also gives a  improvement of our harmonic approximation near the critical current.

\section{ discussion}

 {

 Based on the conventional understanding of von Neumann  measurement, the $\sigma_z$ coupling between qubit and dcSQUID will project the qubit energy state stochastically into either $\left|R \right\rangle $ or $\left|L\right\rangle $  and thus show switching currents with two main current distributions centered at $I_R$ and $I_L$ respectively, if each distribution itself is not too broad to make such distinction. Actually, the experimental results give a different answer: after the NTT group  successfully improved the resolution of the switching current of the dcSQUID~\cite{NTT}, {their raw data of switching currents for flux qubit being in its ground state clearly shows one single peak of the current distribution which continuously shifts from $I_R$ to $I_L $ as the flux dependent behabior of average switcing curent; the existence of such itermediate distribution staying between  $I_R$ and $I_L$ gives a direct evidence of the entanglement between dcSQUID wave function and qubit energy state, otherwise the entanglement of dcSQUID to the qubit flux state should give two current distribuions respectively at $I_R$ and $I_L$ after measurement. Furthermore,} they also do the simulation for a reasonable coupling strength~\cite{NTTT}, and give the same result, namely a single-peak switching current distribution, for the measurement of qubit in the ground state under various external fluxes $\Phi_{q}$. We give a analytical way to understand their result by considering the eigenstates of the qubit-dcSQUID composite system and systematically simplifying the current switching measurement by two-chanel escape dynamics. In sum we give a clear picture to understand why the current switching measurement of flux qubit is a process of wave collapsing into energy eigen states $\left|+ \right\rangle$ and $\left|- \right\rangle $ instead of a process of wave collapsing into flux states $\left|R \right\rangle$ and $\left|L \right\rangle$~\cite{NTTT1}; the more detail discussion is given in the next paragraph.
 
  According to the above analysis which gives Eq.(\ref{eqn:KERH}), 
  once we figure out the appropriate new spin basis to effectively diagonalize the coupled Hamiltonian into two escape chanels by neglecting the minimized tunneling energy between these channels, we can describe the escape dynamics of the qubit-dcSQUID coupled system by two-channel escape dynamics without considering the tunneling between them. Therefore, we can  consider the ground state wave functions of two escape chanels $\left| \psi_+\right\rangle\left| +\right \rangle  $ and $\left| \psi_-\right\rangle\left| -\right \rangle $,  where $\left| \psi_+\right\rangle $ and $\left| \psi_-\right\rangle $ are the corresponding dcSQUID  wave functions, as the eigenstates of the composite system,  and any state $\left| \psi\right\rangle$ of our coupled system with the condition that  the dcSQUID stays in its ground state can be decomposed into these eigen states, which gives 
 \begin{eqnarray}
 \left| \psi\right\rangle=a^+ \left| \psi_+\right\rangle \left| +\right \rangle+ a^- \left| \psi_-\right\rangle\left| -\right \rangle \label{eqn:state}
 \end{eqnarray}
  with $\left| a^-\right|^2 +  \left| a^+\right| ^2=1$; the corresponding density matrix elements of the qubit are $\sigma_{++}\left(0 \right) =\left| a^+\right| ^2$, $\sigma_{--}\left(0 \right)=1-\sigma_{++}\left(0 \right)$, and $\left| \sigma_{-+}\left(0 \right)\right| =\sqrt{\sigma_{++}\left(0 \right)\sigma_{--}\left(0 \right)} $ respectively. Besides, for given SQUID state $ \left| \psi_+\right\rangle$  and $ \left| \psi_-\right\rangle$ we have the corresponding switching probabilities $P_+\left( I_{t+\tau}  \right) $ and $P_-\left(I_{t+\tau} \right) $  for the bias current increasing to $I_{t+\tau}$ within the time period $\left( t,t+\tau\right)  $, which is evaluated by Eq.(\ref{eqn:swprob}) and then is dependent on the current increaing rate $dy/dt$. As a result, the probability $P$ of the dcSQUID switching is given by the distribution 
 \begin{eqnarray}
 P\left( I_{t+\tau} \right)=\sigma_{++}\left(t \right)P_+\left( I_{t+\tau}\right)+\sigma_{--}\left(t \right)P_-\left( I_{t+\tau}\right) 
 \end{eqnarray}
 On the contrary, if the dcSQUID does not switch to the voltage mode during the measurement time $\tau $, we can consider it as a measurement with negative outcome which happens with probability  $1-P\left( I_{t+\tau} \right)$ instead. Consequently, after a measurement with negative outcome we still can get useful knowledge about the system; the diagonal part of density matrix $\sigma_{++}$, $\sigma_{--}$  should change accordingly based on the standard Bayes formula for a posteriori probability, which is simialr to the analysis by Korotkov~\cite{Korotkov}: 
  \begin{eqnarray}
  \sigma_{++}\left(t+\tau \right)= \sigma_{++}\left(t \right) \left( 1-P_+\left( I_{t+\tau} \right) \right)\left\lbrace 1-P(I_{t+\tau})\right\rbrace ^{-1}\nonumber
  \end{eqnarray}
  \begin{eqnarray}
  \sigma_{--}\left( t+\tau\right) =1-\sigma_{++}\left(t+ \tau\right) \label{eqn:DD} 
  \end{eqnarray}
  But once the dcSQUID switches to voltage mode at $I_{t+\tau}$ the density matrix should follow another formula instead:
   \begin{eqnarray}
    \sigma_{++}\left(t+\tau \right)= \sigma_{++}\left(t \right)  P_+\left( I_{t+\tau} \right) \left\lbrace P(I_{t+\tau})\right\rbrace ^{-1} \nonumber \\
    \sigma_{--}\left(t+\tau \right)= \sigma_{--}\left(t \right)  P_+\left( I_{t+\tau} \right) \left\lbrace P(I_{t+\tau})\right\rbrace ^{-1} \label{eqn:DD1} 
    \end{eqnarray}
   
  Because there is no tunneling energy between the two escape channels, {if we assume our measuremnt is an ideal one with no decoherence we can follow the same procedure as in Korotkov~\cite{Korotkov} to give the formula of $ \sigma_{+-}\left(\tau \right)$ ~\cite{decoh}:}
  \begin{eqnarray}
  \left| \sigma_{+-}\left(t+\tau \right)\right| =\sqrt{\sigma_{++}\left( t+\tau\right)\sigma_{--}\left( t+\tau\right)  }.
  \end{eqnarray}
  Then we find
  \begin{eqnarray}
    &\ &\sigma_{+-}\left(t+\tau \right)=\nonumber\\
    &\ &\sigma_{+-}\left(t \right)\exp \left( \frac{i\varepsilon_{+-}\tau}{\hbar } \right)  \sqrt{\frac{\sigma_{++}\left( t+\tau\right)\sigma_{--}\left(t+ \tau\right)}{\sigma_{++}\left( t\right)\sigma_{--}\left( t\right)}  }. \label{eqn:offDD}
    \end{eqnarray}
    Here $\varepsilon_{+-}$ is the energy difference between two states  $ \left| \psi_+\right\rangle \left| +\right \rangle$ and $ \left| \psi_-\right\rangle \left| -\right \rangle$.
  In principle, the equations (\ref{eqn:DD}), (\ref{eqn:DD1}), and (\ref{eqn:offDD}) describe how the density matrix of qubit evolves before the dcSQUID switches to the voltage mode at switching current $I_{SW}$. Obviously, the density matrix will be different at different $I_{SW}$. Moreover, even for the same $I_{SW}$ at different runs the density matrix can be different because the current-increasing histories are not eactly the same due to the presence of the current noise~\cite{current}. 

  Because of the inherent stochastic property of quantum tunnelling, the switching probability has a certain width $\Delta I_{sw}$ such that the displacement($\delta \bar{I}_{sw}$) between two peaks of switching current distribution of qubit states $\left|- \right\rangle $ and $\left|+ \right\rangle $ may be relatively small in comparison. Therefore we need to repeat the  experiment tens of thousands of times to get the change of average value and read out the qubit's state; a single measurement is not enough to get any useful information about the qubit.  This situation( with $\frac{\delta \bar{I}_{sw}}{\Delta I_{sw}}\ll 1$) is similar to the "weak measurement" scheme of Albert et al.~\cite{Albert}(though no post selection step is made in our measurement) and hereafter we call it the "weak measurement limit". In addition, based on the Equations (\ref{eqn:DD}), (\ref{eqn:DD1}), and (\ref{eqn:offDD}) in the "weak measurement limit",  the density matrix  will not show any clear change from its initial value due to the strong overlap between $P_+\left( I \right)$ and $P_-\left( I \right)$. On the contrary, the "von Neumann limit" requires the condition $\frac{\delta \bar{I}_{sw}}{\Delta I_{sw}}\gg 1$. Although a single measurement is still not enough to uniquely characterize the original state of the qubit,  at least we can tell, in this "von Neumann limit", into which energy state does the qubit collapse  after each measurement, which means that our measurement is a projective measurement into $\left|- \right\rangle $ and $\left|+ \right\rangle $. Correspondingly, the density matrix governed by our equations shows good localization in one or other of the two states after the current switching measurement. { In the more general case, while no complete collapse takes place due to the stong overlap between the two switching-current distributions, at least  we can analyze the qubit density matrix according to Eq.(\ref{eqn:DD}), (\ref{eqn:DD1}), and (\ref{eqn:offDD}) to give probabilities of two states after the current switching.}
  
In the Rabi experiments~\cite{Mooij2}, because  $a^+$ and $a_-$ in Eq.(\ref{eqn:state}) should oscillate between 0 and 1 alternatively by applying microwave of correct frequency that is equal to the energy difference between the two states  $ \left| \psi_+\right\rangle \left| +\right \rangle$ and $ \left| \psi_-\right\rangle \left| -\right \rangle$, the total switching probability  $P=\left|a^+ \right|^2P_+ + \left|a^- \right|^2P_- $  should oscillate between $P_+$ and $P_-$. Usually, experimentalists pick up the current at which the difference between $P_+$ and $P_-$ is maximized, therefore, the Rabi or  Ramsey diagram should oscillate with the amplitude $\left| P_+-P_-\right| $. In essence, the analysis of the qubit density matrix is the same as before except that we now measure the switching probability at a given bias current, and once the dcSQUID switches to voltage mode the Eq.(\ref{eqn:DD1}) should apply to the density matrix. Moreover, the larger $\left| P_+-P_-\right| $ can make the density matrix  more localized onto either of two states.

  By considering Eq.(\ref{eqn:GammaDiff}), we find that the negligible decay rate difference at small bias current  makes the difference between the wave functions of the two channels too small to have a significant entanglement between the  qubit and the dcSQUID until the bias current is near the critical current. In addition, in order to infer  the entanglement from the switching current distribution we require a large coupling between the qubit and the dcSQUID to separate the two current distributions corresponding to $\left|\psi_+ \right\rangle $ and $\left|\psi_- \right\rangle $(similar to FIG.\ref{fig:f01}).

     Because the mutual inductance between the qubit and the dcSQUID always exists by experimental design, we have had to consider the qubit and the dcSQUID together as a  quantum system where the coupled hamiltonian governs the energy levels evolving adiabatically with bias current. Therefore, at first glance, we may think the measurement mechanism of qubit-dcSQUID coupled system is different from the standard weak-measurement assumption that the interaction between system and aparatus only turns on at measuring time. But if we regard the fact that in the  weak-coupling limit the qubit and the dcSQUID are not entangled before $I_b$ approaches a certain value near critical current~\cite{NTTT}, it will be more appropriate to think of using the development of a certain degree of  entanglement between the system and apparatus as defining the  time at which the interaction is "turned on".

     The conclusion that we obtain two switching current distributions reflecting  two qubit energy states  is correct only when the decoherence effect from the measurement is  small(in comparison with $\Delta$~\cite{NTTT} ) such that the behavior of the eigen states of the coupled Hamiltonian in Eq.(\ref{eqn:totH}) is good enough to describe the escape dynamics in the measurement process; this situation is similar to the "Hamiltonian-dominated regime" in the review of~\cite{engineer}, if we consider the qubit-dcSQUID together as a quantum system.  The opposite limit is the decoherence-dominated regime in which the two parallel switching current distributions correspond to two qubit "flux" states instead~\cite{NTT2} and our coupled Hamiltonian can not give a complete picture to catch the real dynamics~\cite{NTTT1}. Usually, in a static experiment the qubit density matrix decoherens to a statistical mixture before the switching event is detedcted~\cite{NTTT}

     Note added: When this manuscript was in the final stages of preparation, we received an updated and expanded version of ref.\cite{NTTT}, which treats some of the same issues which are discussed above. While we do not disagree with any of the conclusions of this paper, we want to emphasize our belief that even in the complete absence of the decoherence on experiment starting from a nontrivial superposition of energy eigenstates will yield a two-peak distribution of switching currents.

\section{ Summary}

 In sum, we have given a systematic prescription, based on a harmonic approximation, to determine the proper basis for analysis of the escape dynamics of the coupled qubit-dcSQUID system. The "two-channel" picture we have given captures, we believe, most of the physics of both the static and dynamic measurements, and gives results consistent with those of the numerical analysis\cite{NTTT} of the NTT group for the decoherence-free case; it lends itself to a natural interpretation of the switching distributions in the language of weak and von Neumann measurements.

  The authors thank Hayato Nakano for fruitful discussions of the experiments and of the numerical analysisof ref.\cite{NTTT} and for sending us the updated version of this reference, and Alec Maassen van den Brink for helpful  discussions of theory. We are particularly grateful to  Adrian Lupascu for a careful reading of the manuscript and many constructive comments. This work was supported in part by the Macarthur Profesorship endowed by the John D.and Catherine T.Macarthur Foundation at the University of Illinois.

\appendix
{
\section{Appendix A}

 To deal with the Hamiltonian given in Eq.(\ref{eqn:APXHH1}), in which $H^H=H^H_d+V^H$ and $V^H  =  - \left( {\varepsilon ^H \left( {R}\right)\sin \chi   + \Delta \cos \chi } \right)\tau _x$, we can treat the last term, that is $V^H$, perturbatively. As we know that the lowest two eigenstates of $H^H_d$ are denoted
 as $\left| 0'_+\right\rangle $ and  $\left| 0'_-\right\rangle $, the corresponding eigenenergies in $H^H_d$ are $E_+$ and $E_-$. Following the steps of perturbation theory, we can calculate  correction to the eigenenergy $E_+$, namely

\begin{eqnarray}
 \Delta E _ +    = \frac{{\left| {\left\langle {0'_ +    } \right|V^H\left| {0'_ -  } \right\rangle } \right|}}{{E_ +  - E_ -   }}^2.\label{eqn:delE}
 \end{eqnarray}
Because the first order term $\left\langle {0'_+ } \right|V^H\left| {0'_+  } \right\rangle $is zero automatically, we keep the second order term here. Besides, according to Eq.(18) and (20) we  have 

 \[E_ +  - E_ -\approx \delta \omega\hbar= 2g \omega\hbar \cos \chi \tan \left[ {\pi f_{SQ} } \right]\left( {1 + \left( {\frac{3}{{R_c^{} }}} \right)^2 } \right)
  \]
On the other hand, we can evaluate $\left\langle {0'_ +} \right|V^H\left| {0'_ - }\right\rangle $ in the harmonic approximation:

$\left\langle {0'_+ } \right|V^H\left| {0'_ -  } \right\rangle $
\begin{eqnarray}
 &=&  - \left[ {\left\langle \psi _{0_ + ^{} } \right|\varepsilon^H \left( R \right)\left| \psi _{0_ - ^{} } \right\rangle \sin \chi  + \left\langle {\psi _{0_ + ^{} }}
 \mathrel{\left | {\vphantom {\psi _{0_ + ^{} } \psi _{0_ - ^{} }}}
 \right. \kern-\nulldelimiterspace}
 {\psi _{0_ - ^{} }} \right\rangle \Delta \cos \chi } \right]\nonumber\\
 &=& \frac{{\Delta \left\langle {\psi _{0_ + ^{} }}
 \mathrel{\left | {\vphantom {\psi _{0_ + ^{} } \psi _{0_ - ^{} }}}
 \right. \kern-\nulldelimiterspace}
 {\psi _{0_ - ^{} }} \right\rangle }}{{\sqrt {\varepsilon _{00}^{H2}  + \Delta ^2 } }}\left[ {\frac{{\left\langle \psi _{0_ + ^{} } \right|\varepsilon^H \left( R \right)\left| \psi _{0_ - ^{} } \right\rangle }}{{\left\langle {\psi _{0_ + ^{} }}
 \mathrel{\left | {\vphantom {\psi _{0_ + ^{} } \psi _{0_ - ^{} }}}
 \right. \kern-\nulldelimiterspace}
 {\psi _{0_ - ^{} }} \right\rangle }} - \varepsilon _{00}^H } \right]\nonumber\\
 &=&\frac{{\Delta \left\langle {\psi _{0_ + ^{} }}
  \mathrel{\left | {\vphantom {\psi _{0_ + ^{} } \psi _{0_ - ^{} }}}
  \right. \kern-\nulldelimiterspace}
  {\psi _{0_ - ^{} }} \right\rangle }}{{\sqrt {\varepsilon _{00}^{H2}  + \Delta ^2 } }} \frac{{gk}}{2}\frac{\hbar }{m}\tan \left[ {\pi f_{SQ} } \right] {\left( {\left( {\frac{1}{{\omega _ + ^{}  + \omega _ - ^{} }}} \right) - \frac{1}{{2\omega }}} \right)} \nonumber
\end{eqnarray}
\[ = \frac{{g^3 }}{{16}}\omega \hbar \tan ^3 \left[ {\pi f_{SQ} } \right]\left\langle {\psi _{0_ + ^{} }}
  \mathrel{\left | {\vphantom {\psi _{0_ + ^{} } \psi _{0_ - ^{} }}}
  \right. \kern-\nulldelimiterspace}
  {\psi _{0_ - ^{} }} \right\rangle \sin \chi  \left( {\cos \chi \left( {1 + \left( {\frac{3}{{R_c^{} }}} \right)^2 } \right)} \right)^2 
\]
(Here we have given only the critical steps).  $\left| \psi _{0_ + ^{} } \right\rangle$ and $\left| \psi _{0_ - ^{} } \right\rangle$ are the corresponding SQUID harmonic (ground state) wavefunctions of $\left| 0'_+\right\rangle $ and  $\left| 0'_-\right\rangle $.  Then $\omega_+$ and $\omega_-$ are the corresponding harmonic frequencies. 
 
 Plugging all the above formulae into Eq.(\ref{eqn:delE}), we get  
 \begin{eqnarray}
 \Delta E _ + &\approx& \frac{{\left| {\left\langle {0'_ +
   } \right|V^H \left| {0'_ -   } \right\rangle }
  \right|^2 }}{{\left[ {2\delta\omega \hbar} \right]}} \nonumber\\
  &\approx& \left(
  { \omega \hbar } \right)\frac{{g^5 }}{{1024}}\tan ^5
   \left[ {\pi f_{SQ} } \right]\cos\chi\sin ^2 2\chi \sim 10^{ - 34} J \nonumber
 \end{eqnarray}
 }


 {
\section{Appendix B}

 In this appendix we give a more detailed study of the splitting of the
 lowest two energy levels of the qubit-dcSQUID system, including the effects of
 the deviation of the SQUID wave functions from the pure harmonic form. While
 this more detailed analysis does not by itself allow us to improve the formulae
 for the escape rates $\Gamma_+$ and $\Gamma_-$ calculated in the main text, it is
 useful in a slightly different context, namely the practical implementation of a
 test of the temporal Bell inequalities\cite{LG} at the macroscopic level\cite{TBI}. Our calculation is based on a perturbative treatment of the complete Hamiltonian in Eq.(\ref{eqn:totH}), and its output will be a more accurate value $\chi_{00}$ of the "spin rotation angle" $\chi$ introduced in Eq.(\ref{eqn:angle}) of the main text.

  } 
  Please note that,  for simplicity, from now on we use $\left| n \right\rangle $ to represent the eigen state of $H_{\rm SQ} $ instead of simple harmonic wave function, i.e. $H_{\rm SQ} \left|   {n}\right\rangle  = E_{n} \left|  n \right\rangle$. 
       { 
       Thus, if we approximate the SQUID potential by Eq.(\ref{eqn:APXU}), then the cubic correction to the simple harmonic wave function is already incorporated in the eigen state here (except that we ignore the small escape energy to simplify the analysis).  We will also require $  E_n  \gg \varepsilon _{mn}$, $\Delta $, where $ \varepsilon _{mn}$ is defined in Eq.(\ref{eqn:elmn}); this condition is satisfied for most experimental setups.} 
       
  We can rewrite our total Hamiltonian in the spin representation:
  \begin{eqnarray}
  H = \left( {\begin{array}{*{20}c}
       {H_{{\rm{SQ}}} \left( x \right) + \varepsilon  + \varepsilon _{{\mathop{\rm int}} } \left( x \right)} & \Delta   \\
       \Delta  & {H_{{\rm{SQ}}} \left( x \right) - \varepsilon  - \varepsilon _{{\mathop{\rm int}} } \left( x \right)}  \\
    \end{array}} \right).
  \end{eqnarray}
   Here $H_{SQ}\left( x \right)$ and $\varepsilon _{{\mathop{\rm int}} }\left( x \right)$ are defined in Eq.(\ref{eqn:HSQ})and Eq.(\ref{eqn:vxint}). The basis for this represention is $ \left| {x,\sigma } \right\rangle$
   , where $ \left| x \right\rangle $ and $ \left| \sigma \right\rangle $ $
   \left({\sigma  \in \left\{ {R,L} \right\}} \right)$
    represent the SQUID phase state and qubit flux state respectively. 
 Furthermore, in terms of the energy representation with $
 \left| n,\sigma  \right\rangle$ as basis
   , we can alternatively represent the total Hamiltonian by 
 {
\begin{eqnarray}
H = \left( {\begin{array}{*{20}c}
   {h_{00} } & {h_{01} } & {h_{02} } & .  \\
   {h_{10} } & {h_{11} } & {h_{12} } & .  \\
   {h_{20} } & {h_{21} } & {h_{22} } & .  \\
   . & . & . & .  \\
\end{array}} \right)\label{eqn:Hnm}
\end{eqnarray}

where 
 $ h_{nn}  = \left( {\begin{array}{*{20}c}
     {E_n  + \varepsilon _{nn} } & \Delta   \\
     \Delta  & {E_n  - \varepsilon _{nn} }  \\
  \end{array}} \right)$
    and $h_{\scriptstyle mn \hfill \atop 
      \scriptstyle \left( {m \ne n} \right) \hfill}  = \left( {\begin{array}{*{20}c}
       {\varepsilon _{mn} } & 0  \\
       0 & { - \varepsilon _{mn} }  \\
    \end{array}} \right)$}
     , where
     \begin{eqnarray}
     \varepsilon _{mn}  \equiv \left\langle m \right|\varepsilon  + \varepsilon _{{\mathop{\rm int}} } \left| n \right\rangle. \label{eqn:elmn}
     \end{eqnarray}
       Each element of the diagonal part of Eq.(\ref{eqn:Hnm}) can be rearranged into the form 
       \begin{eqnarray}
       h_{nn}  = E_n  + \sqrt {\varepsilon _{nn} ^2  + \Delta ^2 }  \cdot \left( {\begin{array}{*{20}c}
                {\cos \chi _{nn} } & {\sin \chi _{nn} }  \\
                {\sin \chi _{nn} } & { - \cos \chi _{nn} }  \\
             \end{array}} \right)
       \end{eqnarray}
       with
        \begin{eqnarray}
       \cos \chi _{nn}  \equiv \frac{{\varepsilon _{nn} }}{{\sqrt {\varepsilon _{nn} ^2  + \Delta ^2 } }} , \ \sin \chi _{nn}  \equiv \frac{-\Delta }{{\sqrt {\varepsilon _{nn} ^2  + \Delta ^2 } }}.\label{eqn:angle1}
       \end{eqnarray}
       
         Next, we can divide the total Hamiltonian into a diagonal part ($H_D$ )  and an off-diagonal part ($H-H_D$)
         
         \begin{eqnarray}
         H &=& \left( {\begin{array}{*{20}c}
            {h_{00} } & 0 & 0 & .  \\
            0 & {h_{11} } & 0 & .  \\
            0 & 0 & {h_{22} } & .  \\
            . & . & . & .  \\
         \end{array}} \right) + \left( {\begin{array}{*{20}c}
            0 & {h_{01} } & {h_{02} } & .  \\
            {h_{10} } & 0 & {h_{12} } & .  \\
            {h_{20} } & {h_{21} } & 0 & .  \\
            . & . & . & .  \\
         \end{array}} \right)\nonumber\\
          &=& H_D^{}  + \left( {H - H_D^{} } \right)
         \end{eqnarray} 

In the diagonal part $H_D$, each element $h_{nn}$(, which is the $2\times 2$ diagonal block in $H$,) can be further diagonalized  into 
\begin{eqnarray}
h'_{nn} & =& \left( {\begin{array}{*{20}c}
   {E_n  + \sqrt {\varepsilon _{nn}^2  + \Delta ^2 } } & 0  \\
   0 & {E_n  - \sqrt {\varepsilon _{nn}^2  + \Delta ^2 } }  \\
\end{array}} \right)\nonumber \\
&=& \left( {\begin{array}{*{20}c}
   {E_{n + } } & 0  \\
   0 & {E_{n - } }  \\
\end{array}} \right) \label{eqn:Hnn}
\end{eqnarray} 
 with an appropriate new spin basis 
$
\left| {n,\sigma '} \right\rangle \left( {\sigma ' \in \left\{ { + , - } \right\}} \right)
$
, where
$
\left\{ \begin{array}{l}
 \left| {n, + } \right\rangle  = \cos {\textstyle{{\chi _{nn} } \over 2}}\left| {n,R} \right\rangle  + \sin {\textstyle{{\chi _{nn} } \over 2}}\left| {n,L} \right\rangle  \\ 
 \left| {n, - } \right\rangle  =  - \sin {\textstyle{{\chi _{nn} } \over 2}}\left| {n,R} \right\rangle  + \cos {\textstyle{{\chi _{nn} } \over 2}}\left| {n,L} \right\rangle  \\ 
 \end{array} \right..
$ In other words, the Hamiltonian $H_D$ can be exactly diagonalized by $
\left| {n,\sigma '} \right\rangle ${ ; $H_D \left| {n,\sigma '} \right\rangle  = E_{n,\sigma '} \left| {n,\sigma '} \right\rangle $. Also, the elements in $\left( {H - H_D^{} } \right) $  can be rewritten as }
\begin{eqnarray}
h'_{ mn }  = \varepsilon _{mn} \left( {\begin{array}{*{20}c}
   {\cos {\textstyle{{\chi _{mm}  + \chi _{nn} } \over 2}}} & { - \sin {\textstyle{{\chi _{mm}  + \chi _{nn} } \over 2}}}  \\
   { - \sin {\textstyle{{\chi _{mm}  + \chi _{nn} } \over 2}}} & { - \cos {\textstyle{{\chi _{mm}  + \chi _{nn} } \over 2}}}  \\
\end{array}} \right)
\end{eqnarray}
 {  in this new basis $   \left| {n,\sigma '} \right\rangle $. }
     Hereafter, instead of using $ \left| {\sigma '} \right\rangle $, we denote $
     \left| \sigma  \right\rangle \left( {\sigma  \in \left\{ { + , - } \right\}} \right)
     $ as our new spin basis  for convenience.
      
   In the following, we will consider the off diagonal part $\left( {H - H_D^{} } \right)$ as a perturbation, and argue that the correction to the energy due to it is much smaller than  any  difference between eigenenergies of $H_D$.  Therefore, we can  neglect its effect on eigenenergies and effectively evaluate the energy spectrum of the coupled system.
   
  Let's consider the perturbation correction for the energy levels of $H_D$ { up to second order. }
\begin{eqnarray}
E_{n\sigma }^{{\rm{new}}}&=&E_{n\sigma }^{}  + \left\langle {n\sigma } \right|\left( {H - H_D } \right)\left| {n\sigma } \right\rangle  \nonumber\\
 &+& \sum\limits_{k\sigma '  \ne n\sigma  } {\frac{{\left| {\left\langle {k\sigma '} \right|\left( {H - H_D } \right)\left| {n\sigma } \right\rangle } \right|^2 }}{{E_{n\sigma }^{}  - E_{k\sigma '}^{} }}} + O\left( {M^3 }\right)
\end{eqnarray}


Here $\sigma'$ and $\sigma$ all stand for the new spin basis of $H_D$, where $ {\sigma ,\sigma'  \in \left\{ { + , - } \right\}}$. Because of the off-diagonal property of $\left( {H - H_D^{} } \right)$  , the first order term is exactly zero, $\left\langle {n\sigma } \right|\left( {H - H_D } \right)\left| {n\sigma } \right\rangle  = 0$, and the second order term with the summation $k=n$ , which implies $\sigma'\neq\sigma$ , also vanishes, $\left\langle {n\sigma' } \right|\left( {H - H_D } \right)\left| {n\sigma } \right\rangle  = 0$.   Then we can further simplify the equation to  

\begin{eqnarray}
E_{n\sigma }^{{\rm{new}}}  = E_{n\sigma }^{}  + \sum\limits_{k \ne n, \sigma'} {\frac{{\left| {\left\langle {k\sigma '} \right|\left( {H - H_D } \right)\left| {n\sigma } \right\rangle } \right|^2 }}{{E_{n\sigma }^{}  - E_{k\sigma '}^{} }}}  + O\left( {M^3 } \right)\label{eqn:Enew0}
\end{eqnarray}


Also, the approximation $ \sum\limits_{k \ne n,\sigma'}\frac{{\left| {\left\langle {k\sigma '} \right|\left( {H - H_D } \right)\left| {n\sigma } \right\rangle } \right|^2 }}{{E_{n\sigma }^{}  - E_{k\sigma '}^{} }} \approx \sum\limits_{k  \ne n  }\frac{{\left| {\varepsilon _{kn} } \right|^2 }}{{E_n^{}  - E_k^{} }}$
 is correct to the second order in $M$
  .
Finally, we obtain the new eigenenergies:
\begin{eqnarray}
\begin{array}{l}
  E_{n \pm }^{{\rm{new}}}  = E_{n \pm }^{}  + \sum\limits_{k \ne n} {\frac{{\left| {\varepsilon _{nk} } \right|^2 }}{{E_n^{}  - E_k^{} }}}  + O\left( {M^3 } \right) \\ \label{eqn:Enew}
 \end{array}
\end{eqnarray}

According to Eq.(\ref{eqn:Enew}), the excitation energies (from $n=0$ to $n=1$) for $\sigma=+$ and $\sigma=-$ channels can be derived.
\begin{widetext}
\begin{eqnarray}
\begin{array}{l}
 \rm{\Delta E_+} 
  = \left( {E_1  - E_0 } \right) + \left( {\sum\limits_{k \ne 1} {\frac{{\left| {\varepsilon _{1k} } \right|^2 }}{{E_1^{}  - E_k^{} }}}  - \sum\limits_{k \ne 0} {\frac{{\left| {\varepsilon _{0k} } \right|^2 }}{{E_0^{}  - E_k^{} }}} } \right) + \left( {\sqrt {\varepsilon _{11}^2  + \Delta ^2 }  - \sqrt {\varepsilon _{00}^2  + \Delta ^2 } } \right)+O\left( {M^3 } \right) \\ 
\rm{\Delta E_-} 
  = \left( {E_1  - E_0 } \right) + \left( {\sum\limits_{k \ne 1} {\frac{{\left| {\varepsilon _{1k} } \right|^2 }}{{E_1^{}  - E_k^{} }}}  - \sum\limits_{k \ne 0} {\frac{{\left| {\varepsilon _{0k} } \right|^2 }}{{E_0^{}  - E_k^{} }}} } \right) - \left( {\sqrt {\varepsilon _{11}^2  + \Delta ^2 }  - \sqrt {\varepsilon _{00}^2  + \Delta ^2 } } \right)+O\left( {M^3 } \right) \\ \label{eqn:MW1}
 \end{array}
\end{eqnarray}
\end{widetext}

 It's easy to see that the first two terms of $\rm \Delta E_+$ are { the same as the corresponding terms of $\rm \Delta E_-$, and the third terms of $\rm \Delta E_+$ and $\rm \Delta E_-$ are just different by a sign.} This is consistent with the property that the states $\left| {0, \pm } \right\rangle $ have the same energy correction $ \sum\limits_{k \ne 0} {\frac{{\left| {\varepsilon _{0k} } \right|^2 }}{{E_0^{}  - E_k^{} }}} $ , and the states  $\left| {1, \pm } \right\rangle $ have the same energy correction  $ \sum\limits_{k \ne 1} {\frac{{\left| {\varepsilon _{1k} } \right|^2 }}{{E_1^{}  - E_k^{} }}} $, which are always true for states on the same energy level n. 
Based on this property, we can easily derive the difference of excited energy between spin $+$ and spin $-$ channels.

\begin{eqnarray}
\begin{array}{l}
 {\rm{\Delta E_+ - \Delta E_-}} = \left( {E_{1 + }  - E_{0 + } } \right) - \left( {E_{1 - }  -
 E_{0 - } } \right) \\
  = \left( {E_{0 + }  - E_{0 - } } \right) - \left( {E_{1 + }  - E_{1 - } } \right) \\ 
  = 2\sqrt {\varepsilon _{00}^2  + \Delta ^2 }  - 2\sqrt {\varepsilon _{11}^2  + \Delta ^2 }  \\  \label{eqn:W12}
 \end{array}
\end{eqnarray}

In general, Eq.(\ref{eqn:W12}) is a very good estimation for the difference of resonance frequencies between two spin channels because we can elimate the perturbation correction from $\left( {H - H_D^{} } \right)$ and only need to consider the energy spectrum of $H_D$.

If we want to get rid of the second term on right side of Eq.(\ref{eqn:MW1}), which is $\sum\limits_{k \ne 1} {\frac{{\left| {\varepsilon _{1k} } \right|^2 }}{{E_1^{}  - E_k^{} }}}  - \sum\limits_{k \ne 0} {\frac{{\left| {\varepsilon _{0k} } \right|^2 }}{{E_0^{}  - E_k^{} }}}$, we have to require it to be much smaller than the last term of the formula, $\left( {\sqrt {\varepsilon _{11}^2  + \Delta ^2 }  - \sqrt {\varepsilon _{00}^2  + \Delta ^2 } } \right)$. Then it gives
 \begin{eqnarray}
 \frac{
 \left| {\sum\limits_{k \ne 1} {\frac{{\left| {\varepsilon _{1k} } \right|^2 }}{{E_1^{}  - E_k^{} }}}  - \sum\limits_{k \ne 0} {\frac{{\left| {\varepsilon _{0k} } \right|^2 }}{{E_0^{}  - E_k^{} }}} } \right|}{\left| {\sqrt {\varepsilon _{11}^2  + \Delta ^2 }  - \sqrt {\varepsilon _{00}^2  + \Delta ^2 } } \right|}  \ll 1 \label{eqn:condition}
 \end{eqnarray}

 Typically, we expect $\left( {E_n  - E_m } \right) \sim E_n  \gg \varepsilon _{mn}  \sim \varepsilon _{nn}$, and we can estimate that $\frac{{\left| {\varepsilon _{kn} } \right|^2 }}{{E_n^{}  - E_k^{} }} \sim \frac{{\left| {\varepsilon _{kn} } \right|^2 }}{{E_n^{} }} \sim \frac{{\left| {\varepsilon _{nn} } \right|^2 }}{{E_n^{} }} \ll \left| {\varepsilon _{nn} } \right|$. Therefore, with appropriate dcSQUID bias current $I_b$  and qubit's bias energy $ \varepsilon$, Eq.(\ref{eqn:condition})  usually can be satisfied such that we can have the simpler formulae,
\begin{eqnarray}
\begin{array}{l}
 {\rm{\Delta E_+}} = \left( {E_1  - E_0 } \right) + \left( {\sqrt {\varepsilon _{11}^2  + \Delta ^2 }  - \sqrt {\varepsilon _{00}^2  + \Delta ^2 } } \right) \\ 
 {\rm{\Delta E_-}} = \left( {E_1  - E_0 } \right) - \left( {\sqrt {\varepsilon _{11}^2  + \Delta ^2 }  - \sqrt {\varepsilon _{00}^2  + \Delta ^2 } } \right). \\ 
 \end{array} \label{eqn:MW12b}
\end{eqnarray}
Here $\left( \sqrt {\varepsilon _{11}^2  + \Delta ^2 }  - \sqrt {\varepsilon _{00}^2  + \Delta ^2 } \right)  $ can be further simplified by~\cite{approx} 
\begin{eqnarray}
\sqrt {\varepsilon _{11}^2  + \Delta ^2 }  - \sqrt {\varepsilon _{00}^2  + \Delta ^2 } 
 &\simeq& {\textstyle{{\varepsilon _{00}^{} \left( {\varepsilon _{11}^{}  - \varepsilon _{00}^{} } \right)} \over {\sqrt {\varepsilon _{00}^2  + \Delta ^2 } }}} \nonumber \\
  &=& \cos \chi _{00} \left( {\varepsilon _{11}^{}  - \varepsilon _{00}^{} } \right)
  \label{eqn:leveldiff}
\end{eqnarray}
Next, to evaluate $\varepsilon_{11}$ and $\varepsilon_{00}$ by Eq.(\ref{eqn:elmn}), we can use the approximate formula in Eq.(\ref{eqn:vint}) to replace  $\varepsilon_{int}$, and use the approximate potential in Eq.(\ref{eqn:APXU}) to find the energy levels $\left| n \right\rangle$. After some calculations with the above elements, we finally get the result in Eq.(\ref{eqn:levels}). Actually, the first term in Eq.(\ref{eqn:levels}) is contributed by the square and cubic terms in Eq.(\ref{eqn:vint}) and can be equivalently derived from the change of energy levels of potential in Eq.(\ref{eqn:APXU}) due to the small variation in parameter $k$ by $\delta k=M\pi\cos \chi _{00}\tan \left[ {\pi f_{SQ} } \right] k${ , and the second term in Eq.(\ref{eqn:levels}) can be understood from the non-vanishing ground state expectation value of linear term in Eq.(\ref{eqn:vint}) due to the anharmonic behavior of SQUID ground state $\left| 0\right\rangle $ (see Appendix C).}
\begin{eqnarray}
\begin{array}{l}
\sqrt {\varepsilon _{11}^2  + \Delta ^2 }  - \sqrt {\varepsilon _{00}^2  + \Delta ^2 } \\
 = {\textstyle{\pi \over 2}} M\omega \hbar \tan \left[ {\pi f_{SQ} } \right]\cos \chi _{00}\left( 1+\left( \frac{3}{R_c}\right)^2 \right)
\end{array} \label{eqn:levels}
\end{eqnarray}
 
{ To justify our harmonic-approximation analysis in the section 3, we need to compare the  frequencies  in Eq.(\ref{eqn:deltaomega}) and (\ref{eqn:omegapm}) with those in Eq.(\ref{eqn:MW12b}) and (\ref{eqn:levels}) respectively and the result shows that they are indeed consistent except for a replacement of $\chi$ by $\chi_{00}$, where $\chi_{00}$ is evaluated by using the dcSQUID ground state wavefunction instead of the simple harmonic ground state for $\chi$. }  
 This slight difference could be more significant when the bias current is appoaching its critical value where the anharmonic effect from the cubic term of dcSQUID potential becomes more important. Therefore, if we want to ignore { the  effect of escape and efficiently diagonalize the whole wave function within the well, instead of a  pure harmonic wave function, the SQUID wavefunction $\left| n \right\rangle $ is a more appropriate basis to start with, and the spin angle $\chi_{nn}$ defined in Eq.(\ref{eqn:angle1}) seems better than $\chi$ in Eq.(\ref{eqn:angle}).}
   Finally, we can improve and simplify our harmonic approximation by replacing it { by} rewriting the coupled Hamiltonain of Eq.(\ref{eqn:APXH}) in terms of the new spin basis defined by $\chi_{00}$ and then keeping the diagonal part only.  { Then it gives  an  equation corresponding to Eq.(\ref{eqn:KERH}), namely}
 \begin{eqnarray}
 H_d  =  {\frac{{ - \hbar ^2 }}{{2m}}\partial _R^2  + v \left(
 R \right) + \left( {\varepsilon  \left( R \right)\cos \chi_{00}
   - \Delta \sin \chi_{00}  } \right)\sigma^{00}_z.
 }\label{eqn:APXHH2}
 \end{eqnarray}
 Here we have defiend $ \sigma _z^{00}  = \cos \chi_{00}  \sigma^{00} _z  - \sin \chi_{00} \sigma^{00}_x$. The new parameters corresponding to Eq.(\ref{eqn:Para}) are
\begin{eqnarray}
\begin{array}{l}
 \overline k _ \pm ^{} =k\left( {1 \pm \pi M\cos \chi_{00}  \tan \left[ {\pi f_{SQ} } \right]\left( {1 + \left( {\frac{3}{{R_c^{} }}} \right)^2 } \right)} \right) \\
 \overline R _{c\pm}  = R_c \left( {1 \pm \left( {\frac{3}{{R_c^{} }}} \right)^2 \pi M\cos \chi_{00}\tan \left[ {\pi f_{SQ} } \right]} \right) \\
 k_0^\pm =\mp \left( \sqrt {\varepsilon _{00}^{2}  + \Delta ^2}+\frac{\pi M}{4}\omega \hbar\cos \chi_{00}  \tan \left[ {\pi f_{SQ} } \right]\right. \\
  \left. \times \left( {1 + \left( {\frac{3}{{R_c^{} }}} \right)^2 } \right) 
\right).\label{eqn:Para1}
 \end{array}
\end{eqnarray}

  Comparing the Eq.(\ref{eqn:Para1}) with Eq.(\ref{eqn:Para}), the only changes are the replacement of $\chi$ by $\chi_{00}$ and the small correction { to} $k^{\pm}_{0}$. Finally, we have the same decay rate formula { as  in Eq.(\ref{eqn:GammaPM0})} except that each parameter is modified by the replacement of $\chi$ { by} $\chi_{00}$.

 { In brief, according our spectrum analysis in this section, once we find the "new spin basis" to diagonalize the $H_{nn}$ in Eq.(\ref{eqn:Hnn})(we called it the "first step" here) we can treat the "new off-diagonal part"($H-H_D$) perturbatively as shown in Eq.(\ref{eqn:Enew0}) and Eq.(\ref{eqn:Enew})( the "second step"). Therefore the only  condition required is that $\omega\hbar\gg\varepsilon^H_{00}\left( \rm{or} \  \varepsilon_{mn}\right) $ if we want the perturbation formula Eq.(\ref{eqn:Enew}) to be accurate to the second order; the assumption of $\omega\hbar\gg\Delta$ seems not necessary. 
 Basically, it may be difficult to diagonalize the terms $\partial_R^2$, $\varepsilon\left( R\right)\sigma_z$, and $\Delta\sigma_x$ simutaneously, but we can instead deal with the terms $\partial_R^2$, $\varepsilon\left( R\right)\sigma_z$(that is called the "diagonal terms" in the flux-state representation) first and then treat $\Delta\sigma_x$( the "off-diagonal term")  perturbatively, which only requires the smallness of the "off-diagonal term" to guarantee the correctness of perturbation method. To generalize this (perturbation) method, we can also deal with our total hamiltonian $H$ in similar way but within "new spin representation" where the purpose of choosing the "new spin basis" is to appropriately divide the whole Hamiltnian $H$ into the "diagonal part" and "off-diagonal part" such that we can minimize the "off-diagonal part" in the "new basis"~\cite{minimize}, that is exactly what we do in the "first step".  Although the way to determine the "new spin basis" here is a little bit different from that used in the harmonic approximation in the section 3, their principal ideas are the same.
 }

\section{Appendix C}

 To analyze the energy levels of the dcSQUID, we start with the approximate Hamiltonian   
\begin{eqnarray}
H = \frac{{P^2 }}{{2m}} + \frac{k}{2}R^2  - \beta R^3 \label{eqn:append}
\end{eqnarray}
 with 
 $\beta \equiv \frac{k}{{2R_c }}$ and $\alpha  \equiv \sqrt {\frac{\hbar }{{2m\omega }}} $
 , and treat the cubic term $ - \beta R^3$  perturbatively. Besides, we use $\left| {n} \right\rangle$ for representing energy state of Hamiltonian in Eq.(\ref{eqn:append}) and  $\left| {n'} \right\rangle$ for corresponding simple harmonic state. Then  the dcSQUID's ground state wave function $ \left| {0} \right\rangle $ can be constructed from the simple harmonic wave function $ \left| {n'} \right\rangle $ by perturbation analysis, 

\begin{eqnarray}
 \left| {0} \right\rangle 
 &=& \left| 0' \right\rangle  - \beta \frac{{\left\langle 1' \right|R^3 \left| 0' \right\rangle }}{{E_{0'}  - E_{1'} }}\left| 1' \right\rangle  - \beta \frac{{\left\langle 3' \right|R^3 \left| 0' \right\rangle }}{{E_{0'}  - E_{3'} }}\left| 3' \right\rangle  \nonumber\\ 
 &=& \left| 0' \right\rangle  + \frac{{3\alpha ^3 \beta }}{{\omega \hbar }}\left| 1' \right\rangle  - \frac{{\sqrt 6 \alpha ^3 \beta }}{{3\omega \hbar }}\left| 3' \right\rangle\nonumber .
\end{eqnarray}

 Similarly, we also have the first excited state
\begin{eqnarray}
 \left| {1} \right\rangle
  = \left| 1' \right\rangle  - \frac{{3\alpha ^3 \beta }}{{\omega \hbar }}\left| 0' \right\rangle  + \frac{{\sqrt {72} \alpha ^3 \beta }}{{\omega \hbar }}\left| 2' \right\rangle  + \frac{{\sqrt {24} \alpha ^3 \beta }}{{3\omega \hbar }}\left| 4' \right\rangle \nonumber ,
\end{eqnarray}
and the second excited state
\begin{eqnarray}
\left| {2} \right\rangle 
  = \left| 2'\right\rangle  - \frac{{\sqrt {72} \alpha ^3 \beta }}{{\omega \hbar }}\left| 1' \right\rangle  + \frac{{\sqrt {243} \alpha ^3 \beta }}{{\omega \hbar }}\left| 3' \right\rangle  + \frac{{\sqrt {60} \alpha ^3 \beta }}{{3\omega \hbar }}\left| 5'\right\rangle. \nonumber
\end{eqnarray}

The expectation value of $R$ in these energy levels is
\begin{eqnarray}
\left\langle {0} \right|R\left| {0} \right\rangle  &=& \frac{{3\alpha ^3 \beta }}{{\omega \hbar }}\left( {\left\langle 1' \right|R\left| 0' \right\rangle  + \left\langle 0' \right|R\left| 1' \right\rangle } \right) = \frac{{6\alpha ^4 \beta }}{{\omega \hbar }},\nonumber\\
\left\langle {1} \right|R\left| {1} \right\rangle & =& \frac{{ - 3\alpha ^3 \beta }}{{\omega \hbar }}\left( {\left\langle 1' \right|R\left| 0' \right\rangle  + \left\langle 0' \right|R\left| 1' \right\rangle } \right) \nonumber\\
&+& \frac{{\sqrt {72} \alpha ^3 \beta }}{{\omega \hbar }}\left( {\left\langle 2' \right|R\left| 1' \right\rangle  + \left\langle 1' \right|R\left| 2' \right\rangle } \right) 
  = \frac{{18\alpha ^4 \beta }}{{\omega \hbar }}, \nonumber \\
\left\langle {2} \right|R\left| {2}\right\rangle 
 &=& \frac{{ - \sqrt {72}\alpha ^3 \beta }}{{\omega \hbar }}\left( {\left\langle 2' \right|R\left| 1' \right\rangle  + \left\langle 1' \right|R\left| 2' \right\rangle } \right)\nonumber \\ 
  & +& \frac{{\sqrt {243} \alpha ^3 \beta }}{{\omega \hbar }}\left( {\left\langle 3' \right|R\left| 2' \right\rangle  + \left\langle 2' \right|R\left| 3' \right\rangle } \right) 
   = \frac{{30\alpha ^4 \beta }}{{\omega \hbar }}. \nonumber
\end{eqnarray}

Therefore, the difference between the two expectation values of the linear term in Eq.(\ref{eqn:vint}) is
\begin{eqnarray}
\begin{array}{l}
\pi M \tan \left[ {\pi f_{SQ} } \right]\left( \frac{{3k}}{{R_c }}\left\langle {1} \right|R\left| {1} \right\rangle  - \frac{{3k}}{{R_c }}\left\langle {0} \right|R\left| {0} \right\rangle \right) \\ 
= \pi M \tan \left[ {\pi f_{SQ} } \right]\frac{{3k}}{{R_c }}\frac{{12\alpha ^4 \beta }}{{\omega \hbar }} 
=\frac{{ \pi M }}{2}\omega \hbar\tan \left[ {\pi f_{SQ} } \right] \left( {\frac{3}{{R_c }}} \right)^2 .
\end{array}\nonumber
\end{eqnarray}
According to the definition of $\varepsilon_{mn}$ in Eq.(\ref{eqn:elmn}), we can insert this result into Eq.(\ref{eqn:leveldiff}), and give the second term in 
Eq.(\ref{eqn:levels}). 
On other hand, we can also calculate all the other terms in 
 $\varepsilon _{00}$ and $\varepsilon _{11}$ by Eq.(\ref{eqn:elmn}), and as a result we have
\[
\begin{array}{l}
 \varepsilon _{00}^{}  = \frac{{\pi M}}{2}\omega \hbar \tan \left[ {\pi f_{SQ} } \right]\left( {\frac{1}{2} + \frac{1}{2}\left( {\frac{3}{{R_c }}} \right)^2 } \right) + \varepsilon  - \pi Mk\tan \left[ {\pi f_{SQ} } \right] \\ 
 \varepsilon _{11}^{}  = \frac{{\pi M}}{2}\omega \hbar \tan \left[ {\pi f_{SQ} } \right]\left( {\frac{3}{2} + \frac{3}{2}\left( {\frac{3}{{R_c }}} \right)^2 } \right) + \varepsilon  - \pi Mk\tan \left[ {\pi f_{SQ} } \right] \\ 
 \end{array}
\]


\begin{thebibliography}{99}
\bibitem{plateau}{ Quantitatively, the trailing plateau is of this order of magnitude, but its actual value depends on the details of the SQUID circuit.}
\bibitem{three}{ We consider the qubit designed with three Josephson junctions, which is discussed in most of papers~\cite{Mooij1,Mooij2}.  The qubit maximum persistent current is given by $I_p \approx 2 \alpha\pi  E_J/\Phi_0$, where $\alpha$ is the area ratio between the (smaller) middle junction and the other outer ones and $E_J$ is the Josephson coupling energy of outer ones. }
\bibitem{turning} { This formula is true for the bias current approaching its critical value. More general expression for the turning point is given by $R_c^2=27\left( 1-\left( \frac{\pi}{2}-x_0\right)\tan \left[ x_0\right]  \right) $. Qualitatively, there would not be a big difference between these two formulas for the turning point if we consider the switching current behavior near the critical value. }
\bibitem{SQUID DOF}{  The other variable of the dcSQUID, which corresponds to the phase difference bwteen the two junctions, is uncontrollable and unmeasurable but can be integrated out by the path integral method, see ref~\cite{NTTT}.} Its effect is independent of $\sigma_z$ and much smaller than the coupling term between the qubit and dc SQUID.
\bibitem{approx}  The approximation in Eq.(\ref{eqn:leveldiff}) is more appropriate for the NTT data where $g\omega \hbar$ can be much smaller than $\sqrt{\varepsilon_{00}^2+\Delta^2} $.
\bibitem{NTTT1} The strong decoherence can make the qubit state into a statistical mixture of flux states ~\cite{NTTT}; the qubit is trapped in either side of its double potential probabilistically due to the strong decoherence which prohibits the tunneling between the wells. This effect is not included in our dynamical analysis of qubit-SQUID composite system.
\bibitem{decoh} { The decoherence during the measurement could reduce the off diagonal density matrix element. Therefore, we have $ \left| \sigma_{+-}\left(t+\tau \right)\right| <\sqrt{\sigma_{++}\left( t+\tau\right)\sigma_{--}\left( t+\tau\right)  }$.}
\bibitem{current} According to Eq.(\ref{eqn:swprob}), $P_+(I)$ and $P_+(I)$ also depend on the current history if the current does not increase steadily.
\bibitem{minimize}  The purpose for minimization is to make the perturbation correction as small as possible. Therefore, roughly speaking, although the motivations are different, the  minimization criterion in  Appendix A can be  Eq.(\ref{eqn:condition}), and  the corresponding one in the  harmonic approximation of section 3 is the requirement that the energy correction to $H_d$(in Eq.(\ref{eqn:KERH}) due to the perturbation $V^H$ should be smaller than the energy scale of the escape rate near the current corresponding to the maximum switching probability.
\end{thebibliography}
\end{document}